\documentclass[preprint,superscriptaddress,nofootinbib,amsmath,amssymb,prd]{revtex4-2}
\usepackage{graphicx}
\usepackage[colorlinks,linkcolor=magenta,anchorcolor=cyan,citecolor=blue]{hyperref}
\usepackage{physics}
\usepackage{float}
\usepackage{subcaption}
\DeclareUnicodeCharacter{2212}{-}
\DeclareUnicodeCharacter{2243}{$\simeq$}
\usepackage{aas_macros}

\usepackage{ulem}

\begin{document}

\title{Explanation of high redshift luminous galaxies from JWST by early dark energy model}

\author{Jun-Qian Jiang}
\affiliation{School of Physical Sciences, University of
Chinese Academy of Sciences, Beijing 100049, China}
\author{Weiyang Liu}%
\affiliation{%
 Key Laboratory of Space Astronomy and Technology, National Astronomical Observatories, Chinese Academy of Sciences, Beijing 100101, China
}%
\affiliation{%
 School of Astronomy and Space Science, University of Chinese Academy of Sciences, Beijing, 100049, China P.R.
}%

\author{Hu Zhan}
\affiliation{%
 Key Laboratory of Space Astronomy and Technology, National Astronomical Observatories, Chinese Academy of Sciences, Beijing 100101, China
}%
\affiliation{%
The Kavli Institute for Astronomy and Astrophysics, Peking University, Beijing 100871, China
}%
\author{Bin Hu}
\affiliation{%
School of Physics and Astronomy, Beijing Normal University, Beijing 100875, China
}%

\begin{abstract}
Recent observations from the James Webb Space Telescope (JWST) have uncovered massive galaxies at high redshifts, with their abundance significantly surpassing expectations. This finding poses a substantial challenge to both galaxy formation models and our understanding of cosmology. Additionally, discrepancies between the Hubble parameter inferred from high-redshift cosmic microwave background (CMB) observations and those derived from low-redshift distance ladder methods have led to what is known as the ``Hubble tension''. Among the most promising solutions to this tension are Early Dark Energy (EDE) models. In this study, we employ an axion-like EDE model in conjunction with a universal Salpeter initial mass function to fit the luminosity function derived from JWST data, as well as other cosmological probes, including the CMB, baryon acoustic oscillations (BAO), and the SH0ES local distance ladder.
Our findings indicate that JWST observations favor a high energy fraction of EDE, \( f_\text{EDE} \sim 0.2 \pm 0.03 \), and a high Hubble constant value of \( H_0 \sim 74.6 \pm 1.2 \) km/s/Mpc, even in the absence of SH0ES data. This suggests that EDE not only addresses the \( H_0 \) tension but also provides a compelling explanation for the observed abundance of massive galaxies identified by JWST.
\end{abstract}

\maketitle
\newpage

\section{Introduction}

Hubble tension is one of the toughest challenges for the $\Lambda$CDM model nowadays.
The measurements of the CMB power spectrum by Planck satellite assuming the $\Lambda$CDM model inferred the Hubble constant ($67.4 \pm 0.5$) km/s/Mpc \cite{Planck:2018vyg}.
However, observations on Cepheid-calibrated SNeIa find $H_0 > 70$ km/s/Mpc.
For example, the SH0ES team reports $73.04 \pm 1.04$ km/s/Mpc \cite{Riess:2021jrx}, which has a $\sim 5\sigma$ tension with the Planck results.
Since this tension is unlike systematical uncertainties or due to individual observation,
modifications to the $\Lambda$CDM model may be required to restore the tension.

Early Dark Energy (EDE)
(see e.g. \cite{Karwal:2016vyq,Poulin:2018cxd,Kaloper:2019lpl,Agrawal:2019lmo,Lin:2019qug,Smith:2019ihp,Niedermann:2019olb,Sakstein:2019fmf,Ye:2020btb,Gogoi:2020qif,Braglia:2020bym,Lin:2020jcb,Seto:2021xua,Nojiri:2021dze,Karwal:2021vpk,Rezazadeh:2022lsf})
is one of the classes of modified cosmological models that promise to resolve Hubble tension.
In such models, there is an ingredient that behaves like dark energy at a very early time,
and then its energy density starts to decay faster than radiation around matter-radiation equality.
While the EDE models are able to raise the Hubble constant inferred from the CMB and are favored by some recent ground-based CMB observations \cite{Chudaykin:2020acu,Chudaykin:2020igl,Jiang:2021bab,LaPosta:2021pgm,Smith:2022hwi,Jiang:2022uyg},
they have encountered a number of challenges, especially conflicts with LSS observations \cite{Hill:2020osr,Ivanov:2020ril,DAmico:2020ods,Vagnozzi:2021gjh,Reeves:2022aoi,Vagnozzi:2023nrq}.

The recent observations by JWST discover massive galaxies at high redshift with number densities far exceeding those predicted by the $\Lambda$CDM model \cite{Labbe:2022ahb,Boylan-Kolchin:2022kae}.
However, it has been shown that EDE models can lead to a higher abundance of halos \cite{Klypin:2020tud,Forconi:2023hsj,Liu:2024yan,Shen:2024hpx}.
In this work, we employ the method in Ref.\cite{Liu:2024yan} and the luminosity function of galaxies based on JWST observations,
which contains a large number of galaxies over a wide redshift range.

A number of possible resolutions to explain the outnumbered abundance of the bright galaxies at $z\geq10$ have been proposed since its observational confirmation by JWST.
For example, one of the straightforward solutions is high star formation efficiency.
In this scenario, the star formation efficiency is assumed to be variable and to have higher values at high redshift.
Some of the hydrodynamical simulations seems to indicate this point (e.g.~\cite{2023A&A...677L...4P,2024A&A...689A.244C}).
This effect may be due to the feedback-free starbursts (e.g.~\cite{2023MNRAS.523.3201D,2024arXiv240618352H}), which is excepted for galaxies with high densities and low metallicities.
At the epoch of pre-reionization, the star formation efficiency is also expected to lack suppression from the UV background radiation (e.g.~\cite{Harikane:2022rqt}).
Another type of solution doubts the UV luminosity.
Ref.~\cite{Ferrara:2022dqw} considered that dust can be efficiently ejected during the very first phases of galaxy build-up and the dust attenuation is negligible, which makes the galaxies brighter at high reshifts (see also~\cite{Iocco:2024rez}).
In addition, AGN activities may also partly contribute to the UV luminosity.
Ref.~\cite{2023ApJ...959...39H} reported some AGNs in galaxy samples, especially CEERS\_1019 and GN-z11~\cite{Bunker:2023lzn,CEERSTeam:2023qgy,2023ApJ...952...74T}.
However, a majority of the galaxies show no signatures of AGNs and may be truly stellar massive~\cite{2023ApJ...959...39H}.
Another approach is motivated by the fact that the typical application of the classical Salpeter Initial Mass Function (IMF) and its descendants (\cite{1955ApJ...121..161S,2019NatAs...3..482K}) assumes that the Galactic IMF is identical to that of the extragalactic systems (\cite{2023ApJ...951L..40S}), which may not stand as firm for the galaxies at high redshift.
On the contrary, for a non-universal IMF, the higher abundance of bright galaxies could be attributed to the higher contribution of the massive stars (top-heavy, \cite{2023ApJ...951L..40S,2024MNRAS.tmp.1525C}).
However, further studies, e.g. Refs.~\cite{2024A&A...686A.138C,2024arXiv240618352H} suggest that it may not be enough to resolve this issue, due to the lack of direct observational evidence, and that the stronger stellar feedback may counteract the top-heavier IMF.
Here, we do not modify the understanding of astrophysics (i.e., maintaining time-independent star formation efficiency and universal IMF),
and study if EDE, a cosmological proposition that appears to have been observationally detected in CMB observation, can better resolve the tension of the galaxy abundance between JWST and previous prediction. 

This paper is organized as follows.
We summarized our methodology and data employed in \autoref{sec:method_data}.
The results are shown in \autoref{sec:results} and we discuss them there.
Finally, we conclude in \autoref{sec:conclusion}.

\section{Methodology and Data} \label{sec:method_data}

The constraints on the Hubble constant derived from the CMB data arise mainly from observations of the acoustic peak separation:
\begin{equation}
    \theta_s = \frac{r_s}{D_A} \propto r_s H_0 \,, 
\end{equation}
where $r_s$ is the sound horizon and $D_A$ is the angular distance to the last scattering surface.
While $\theta_s$ is well constrained by current CMB observations,
a higher $H_0$ means a lower $r_s$ unless modifications to the evolution of the late Universe alter the relation between $D_A$ and $H_0$.
However, the latter possibility is tightly constrained by observations of BAO and SNeIa, which means that $r_s$ needs to be reduced \cite{Bernal:2016gxb,Aylor:2018drw,Knox:2019rjx,Jiang:2024xnu}. 
In the EDE models, extra energy injections before recombination, which lead to a faster expansion of the Universe at that time,
reduce the sound horizon $r_s$.
After that, the energy density of the EDE decays faster than the radiation to avoid breaking the fitting to the CMB.
There are many implementations of the EDE model.
For simplicity, we consider the original axion-like EDE model \cite{Poulin:2018cxd}.
In this model, the EDE is realized by a canonical scalar field with the following potential:
\begin{equation}
    V( \phi ) = m^2 f^2 ( 1-\cos\theta)^{n} \text{, where } \theta=\phi /f \in [-\pi, \pi]
\end{equation}
The field is frozen initially due to the Hubble friction.
As the expansion rate of the Universe decreases, it will roll down and oscillate at the bottom of the potential.
In this period, the energy density of the EDE decays with the equation of state $w = (n-1)/(n+1)$.
It has been shown $n \sim 3$ is favored by observations \cite{Poulin:2018cxd}.
Therefore, we fix $n=3$ in this work.

The luminosity function $\dd n / \dd M_\text{UV}$ is the number density of galaxies with respect to their absolute magnitude $M_\text{UV}$ at the ultraviolet band.
Our modeling of the luminosity function is shown in Ref.\cite{Lin:2019qug}.
Here we briefly provide the key points.
The luminosity function can be decomposed into three components:
\begin{equation}
    \dv{n}{M_\text{UV}} = \dv{n}{M_h} \dv{M_h}{M_*} \dv{M_*}{M_\text{UV}} \,,
\end{equation}
where $M_h$ is the halo mass and $M_*$ is the stellar mass.
The first part is the halo mass function.
We use the Sheth-Mo-Tormen fitting function \cite{Sheth:1999su}:
\begin{equation} \label{eq:SMT}
    f_{\rm SMT}(\sigma) = A\sqrt{2a/\pi}\nu\exp(-a\nu^2/2)(1+(a\nu^2)^{-p}) \,,
\end{equation}
where $\sigma$ is the standard deviation of the fluctuation of the density in a sphere containing matter with mass $M_\text{m}=M_h$ and $\nu = \delta_\text{crit} / \sigma$.
It contains three parameters \{$A, a, p$\}.
As for $\dd M_h / \dd M_*$,
instead of using a constant linear stellar-to-halo mass ratio $\epsilon$ (a.k.a. mass-independent and redshift-independent star formation efficiency),
we utilize the redshift-independent but mass-dependent formula in Ref.\cite{2021ApJ...922...29S}:
\begin{equation} \label{eq:galaxy_formation_efficiency}
    \frac{M_*}{M_h} = 2N \left[ \left( \frac{M_h}{M_c} \right)^{-\beta} + \left( \frac{M_h}{M_c} \right)^{-\gamma} \right]^{-1} \,,
\end{equation}
    where $N$ is an overall amplitude factor and $M_c$ is the characteristic halo mass.
Since very rare AGNs were detected at $z \geq 8$ \cite{CEERSTeam:2023qgy,2023arXiv230609142S},
we extrapolate an AGN-free stellar-to-halo mass ratio to $z=12$ by using $\gamma = 0.01$.
We consider a scaling relation between the
halo mass $M_\ast$
and the luminosity of a galaxy at the UV band $M_{UV}^\text{dust-free}$
\begin{equation}
    M_\text{UV}^\text{dust-free} = [\lg(M_*) - a_0] / a_1 \,.
\end{equation}
For consistency, we use the coefficients $a_0, a_1$ given in Ref.\cite{2021ApJ...922...29S}, based on the fitting till $z \sim 10$.
Finally, we take the dust attenuation rule \cite{Meurer:1999jj} to account for dust absorption in the UV band and re-emission in the IR band:
\begin{equation}
    A_{1600} = 4.43 + 1.99 \beta \;,
\end{equation}
where $A_{1600}$ is the dust attenuation factor at $1600$ \AA \, and $\beta = -0.17 M_\text{UV}^\text{dust-free} - 5.40$ \cite{2023MNRAS.520...14C}.
The observed $M_\text{UV}$ is the sum of the dust-free $M_\text{UV}^\text{dust-free}$ and the dust attenuation factor:
\begin{equation}
    M_\text{UV} = M_\text{UV}^\text{dust-free} + A_{1600} \, .
\end{equation}

We take the luminosity function data observed by JWST in Refs.\cite{Bouwens:2022gqg,2023MNRAS.523.1036B,Castellano:2022ikm,2023MNRAS.518.6011D,2022ApJ...940L..55F,Harikane:2022rqt,2024ApJ...960...56H,2024ApJ...960...56H,2023ApJ...946L..35M,Perez-Gonzalez:2023wta} and only use data points at $z\in [8,\, 12]$ due to the high photometric redshift uncertainty at higher redshift.
The volume differences caused by differences in the distance-redshift relation of different cosmologies are considered.
We use the CMB data from Planck 2018 \cite{Planck:2018vyg}, containing TT,TE,EE spectrum, and lensing spectrum.
Meanwhile, we include the BAO data from 6dF \cite{Beutler:2011hx}, SDSS DR7 MGS \cite{Ross:2014qpa}, BOSS DR12 \cite{BOSS:2016wmc}, eBOSS DR16 \cite{eBOSS:2020yzd}, which contains measuments based on observations on ELG, LRG, QSO, Ly-$\alpha$ auto-correlation and its cross-correlation with QSO.
SNeIa data from the Pantheon+ analysis \cite{Brout:2022vxf} is also included.
In additionally, we also consider the case with the SH0ES data \cite{Riess:2021jrx} included, which we use the full covariance matrix with Pantheon+.
Recently, the DESI team reported the measurement of BAO from their year of observations \cite{DESI:2024mwx}.
We show the impact of their BAO measurements in \autoref{sec:DESI}.

\begin{table}[h]
    \centering
    \begin{tabular}{|c|c|} \hline
        parameter           & prior \\ \hline
        $n_s$               & $\mathcal{U}[0.8, 1.2]$ \\
        $H_0$               & $\mathcal{U}[60, 80]$ \\
        $\log(10^{10}) A_s$ & $\mathcal{U}[1.61, 3.91]$ \\
        $\omega_b$          & $\mathcal{U}[0.01, 0.03]$ \\
        $\omega_\text{cdm}$ & $\mathcal{U}[0.1, 0.15]$ \\
        $\tau_\text{reio}$  & $\mathcal{U}[0.01, 0.8]$ \\ \hline
        $\ln z_c$           & $\mathcal{U}[7.5, 9]$ \\
        $f_\text{EDE}$      & $\mathcal{U}[0, 0.3]$ \\
        $\Theta_\text{ini}$ & $\mathcal{U}[0, 3.1]$ \\ \hline
        $A$                 & $\mathcal{N}(0.3295, 0.0003)$ \\
        $a$                 & $\mathcal{N}(0.7689, 0.0011)$ \\
        $p$                 & $\mathcal{N}(0.2536, 0.0026)$ \\
        $\log_{10}M_c$      & $\mathcal{N}(11.5, 0.2)$ \\
        $N$                 & $\mathcal{N}(0.0297, 0.0065)$ \\
        $\beta$             & $\mathcal{N}(1.35, 0.26)$ \\ \hline
    \end{tabular}
    \caption{Parameters and their priors used in our model, see \autoref{sec:method_data} for the meaning of these parameters. $\mathcal{U}$ means uniform distribution and $\mathcal{N}(\nu, \sigma)$ means Gaussian distribution.}
    \label{tab:prior}
\end{table}

To perform the joint Bayesian analysis, we utilized \texttt{cobaya} \cite{Torrado:2020dgo} for Markov Chain Monte Carlo (MCMC) sampling.
All the chains have reached Gelman-Rubin criterion $R-1<0.03$.
Besides the six parameters \{$n_s, H_0, A_s, \omega_\text{b}, \omega_\text{cdm}, \tau_\text{reio}$\} for $\Lambda$CDM models,
we sample three phenomenological parameters for the axion-like EDE model:
the redshift $z_c$ when the EDE reached its maximum energy fraction, its energy fraction $f_\text{EDE}$ at that time,
and the initial position of the axion-like EDE field $\Theta_\text{ini}$.
The parameters for modeling the luminosity function mentioned above are also sampled.
All the parameters for our models and their priors are summarized in \autoref{tab:prior}.
We use \texttt{BOBYQA} \cite{2018arXiv180400154C,2018arXiv181211343C} to find the best-fit points.

\section{Results and Discussion} \label{sec:results}

\begin{figure}[H]
    \centering
    \includegraphics[width=\textwidth]{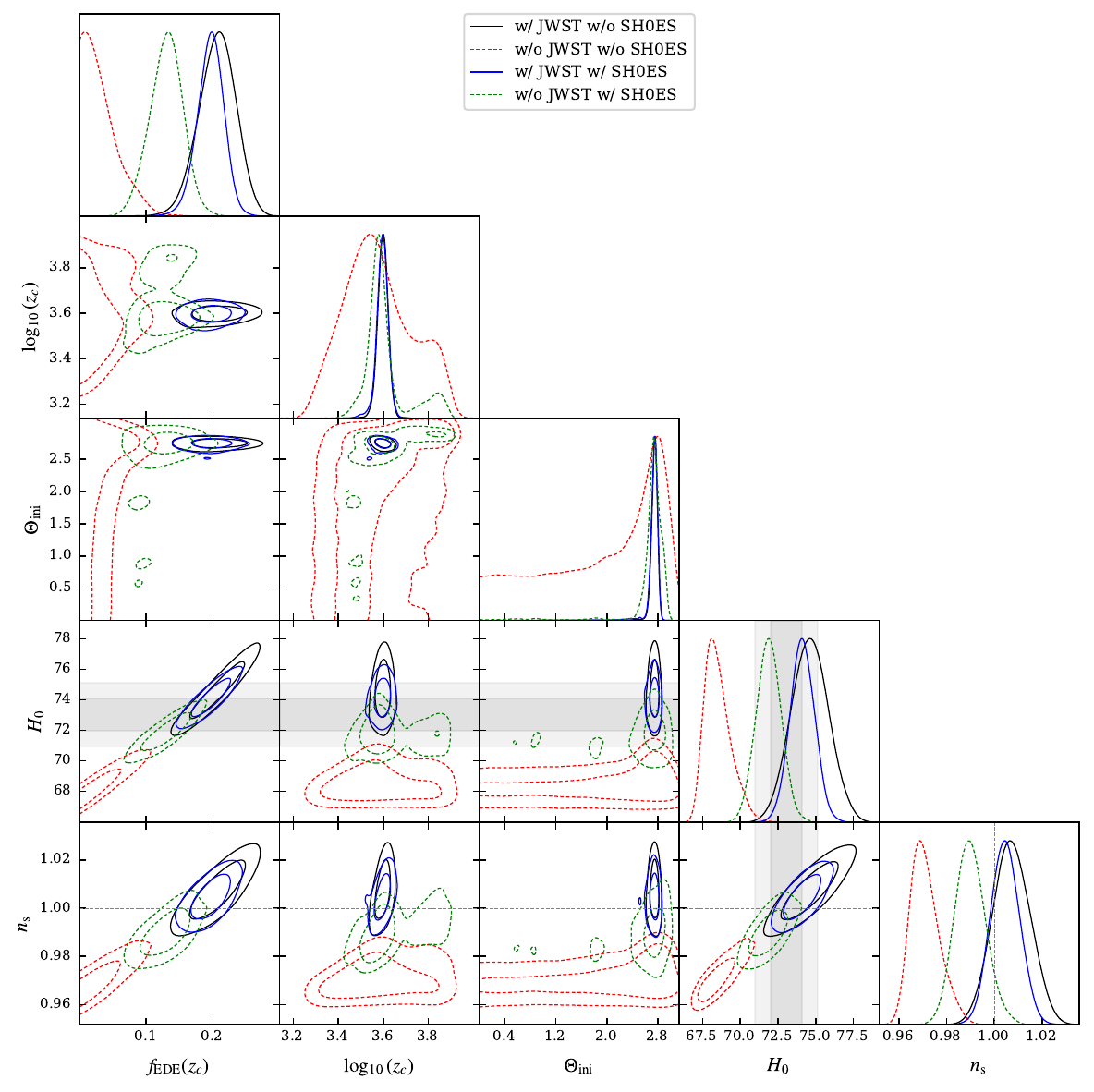}
    \caption{Marginalized posterior distributions for important parameters. The contours show the 68\% and 95\% confidence levels. We also show the SH0ES \cite{Riess:2021jrx} constraints on $H_0$ in gray bands.}
    \label{fig:triangle}
\end{figure}

\begin{table}[]
    \centering\scriptsize
\begin{tabular} {| l | c| c| c| c|}
\hline\hline
 & axionEDE (w/ JWST) & axionEDE (w/o JWST) & $\Lambda$CDM (w/ JWST) & $\Lambda$CDM (w/o JWST)\\
\hline
$\ln(z_c)                  $ & $8.284(8.238)\pm 0.054     $ & $8.24(8.120)^{+0.29}_{-0.46}$ &                              &                             \\
$f_\mathrm{EDE}(z_c)       $ & $0.206(0.2154)^{+0.029}_{-0.026}$ & $< 0.0412(0.0154)          $ &                              &                             \\
$\Theta_\mathrm{ini}       $ & $2.743(2.7734)^{+0.055}_{-0.034}$ & $1.94(2.25)^{+1.2}_{-0.44} $ &                              &                             \\
\hline
$H_0                       $ & $74.6(75.11)\pm 1.2        $ & $68.51(67.85)^{+0.59}_{-1.0}$ & $67.79(67.659)\pm 0.39     $ & $67.88(68.072)\pm 0.40     $\\
$n_\mathrm{s}              $ & $1.0071(1.0077)\pm 0.0078  $ & $0.9710(0.9684)^{+0.0046}_{-0.0071}$ & $0.9674(0.96700)\pm 0.0036 $ & $0.9645(0.96616)\pm 0.0037 $\\
$\Omega_\mathrm{b} h^2     $ & $0.02295(0.022855)\pm 0.00022$ & $0.02252(0.022459)^{+0.00016}_{-0.00020}$ & $0.02236(0.022321)\pm 0.00014$ & $0.02236(0.022384)\pm 0.00014$\\
$\Omega_\mathrm{c} h^2     $ & $0.1450(0.14618)\pm 0.0044 $ & $0.1232(0.12145)^{+0.0017}_{-0.0035}$ & $0.12028(0.12051)\pm 0.00087$ & $0.11997(0.11953)\pm 0.00088$\\
$\log(10^{10} A_\mathrm{s})$ & $3.097(3.0870)\pm 0.015    $ & $3.051(3.0563)\pm 0.015    $ & $3.062(3.0550)^{+0.014}_{-0.016}$ & $3.040(3.0468)\pm 0.014    $\\
$\tau_\mathrm{reio}        $ & $0.0613(0.0552)^{+0.0069}_{-0.0081}$ & $0.0552(0.0580)\pm 0.0073  $ & $0.0624(0.0590)^{+0.0070}_{-0.0081}$ & $0.0524(0.0574)\pm 0.0071  $\\
\hline
$A                         $ & $0.32949(0.329504)\pm 0.00030$ &                              & $0.32949(0.329517)\pm 0.00030$ &                             \\
$a                         $ & $0.7688(0.76886)\pm 0.0011 $ &                              & $0.7687(0.76870)\pm 0.0011 $ &                             \\
$p                         $ & $0.2537(0.25334)\pm 0.0026 $ &                              & $0.2537(0.25389)\pm 0.0026 $ &                             \\
$\log_{10} M_c             $ & $11.52(11.465)^{+0.11}_{-0.13}$ &                              & $11.192(11.158)^{+0.085}_{-0.10}$ &                             \\
$N                         $ & $0.0188(0.01729)^{+0.0025}_{-0.0046}$ &                              & $0.0249(0.02354)^{+0.0029}_{-0.0042}$ &                             \\
$\beta                     $ & $1.456(1.483)^{+0.077}_{-0.093}$ &                              & $1.509(1.517)^{+0.080}_{-0.098}$ &                             \\
\hline
$ \chi^2 $ & $4437.8$ & $4189.1$ & $4482.5$ & $4189.4$ \\ \hline
$\chi^2 - \chi^2_\text{$\Lambda$CDM}$ & $-44.8$ & $-0.3$ & & \\ \hline \hline
\end{tabular}
    \caption{Marginalized posterior on model parameters where SH0ES data is not included. We show the mean values, 68\% confidence levels and put the bestfit values in the parentheses.}
    \label{tab:posterior}
\end{table}

\begin{table}[]
    \centering \scriptsize
\begin{tabular} {| l | c| c| c| c|}
\hline\hline
 & axionEDE (w/ JWST) & axionEDE (w/o JWST) & $\Lambda$CDM (w/ JWST) & $\Lambda$CDM (w/o JWST)\\
\hline
$\ln(z_c)                  $ & $8.279(8.291)\pm 0.063     $ & $8.317(8.259)^{+0.047}_{-0.20}$ &                              &                             \\
$f_\mathrm{EDE}(z_c)       $ & $0.196(0.2107)^{+0.021}_{-0.018}$ & $0.130(0.1326)^{+0.026}_{-0.023}$ &                              &                             \\
$\Theta_\mathrm{ini}       $ & $2.734(2.7573)^{+0.066}_{-0.031}$ & $2.67(2.742)^{+0.20}_{-0.025}$ &                              &                             \\
\hline
$H_0                       $ & $74.11(74.71)\pm 0.83      $ & $71.89(71.98)\pm 0.83      $ & $68.60(68.714)\pm 0.38     $ & $68.66(68.558)\pm 0.38     $\\
$n_\mathrm{s}              $ & $1.0045(1.0102)\pm 0.0060  $ & $0.9899(0.9912)^{+0.0059}_{-0.0068}$ & $0.9716(0.97423)\pm 0.0036 $ & $0.9684(0.96589)\pm 0.0035 $\\
$\Omega_\mathrm{b} h^2     $ & $0.02292(0.022967)\pm 0.00022$ & $0.02283(0.022797)^{+0.00020}_{-0.00023}$ & $0.02253(0.022598)\pm 0.00013$ & $0.02253(0.022539)\pm 0.00013$\\
$\Omega_\mathrm{c} h^2     $ & $0.1435(0.14571)\pm 0.0034 $ & $0.1333(0.13371)^{+0.0032}_{-0.0036}$ & $0.11865(0.11852)\pm 0.00083$ & $0.11840(0.11872)\pm 0.00084$\\
$\log(10^{10} A_\mathrm{s})$ & $3.094(3.0970)\pm 0.015    $ & $3.069(3.0752)\pm 0.014    $ & $3.072(3.0875)^{+0.015}_{-0.017}$ & $3.047(3.0402)\pm 0.015    $\\
$\tau_\mathrm{reio}        $ & $0.0607(0.0610)^{+0.0070}_{-0.0079}$ & $0.0564(0.0594)\pm 0.0070  $ & $0.0686(0.0766)^{+0.0075}_{-0.0090}$ & $0.0572(0.0533)^{+0.0067}_{-0.0075}$\\
\hline
$A                         $ & $0.32949(0.329470)\pm 0.00030$ &                              & $0.32949(0.329476)\pm 0.00030$ &                             \\
$a                         $ & $0.7688(0.76873)\pm 0.0011 $ &                              & $0.7687(0.76865)\pm 0.0011 $ &                             \\
$p                         $ & $0.2537(0.25362)\pm 0.0026 $ &                              & $0.2537(0.25383)\pm 0.0026 $ &                             \\
$\log_{10} M_c             $ & $11.50(11.516)^{+0.10}_{-0.13}$ &                              & $11.181(11.186)^{+0.085}_{-0.099}$ &                             \\
$N                         $ & $0.0191(0.01695)^{+0.0024}_{-0.0045}$ &                              & $0.0254(0.02273)^{+0.0028}_{-0.0043}$ &                             \\
$\beta                     $ & $1.460(1.453)^{+0.078}_{-0.093}$ &                              & $1.514(1.515)^{+0.084}_{-0.096}$ &                             \\
\hline
$ \chi^2 $ & $4487.3$ & $4242.7$ & $4564.6$ & $4271.3$ \\ \hline
$\chi^2 - \chi^2_\text{$\Lambda$CDM}$ & $-77.3$ & $-28.6$ & & \\ \hline \hline
\end{tabular}
    \caption{Marginalized posterior on model parameters where SH0ES data is included. We show the mean values, 68\% confidence levels and put the bestfit values in the parentheses.}
    \label{tab:posterior_wH0}
\end{table}

\begin{figure}
    \centering
    \includegraphics[width=\linewidth]{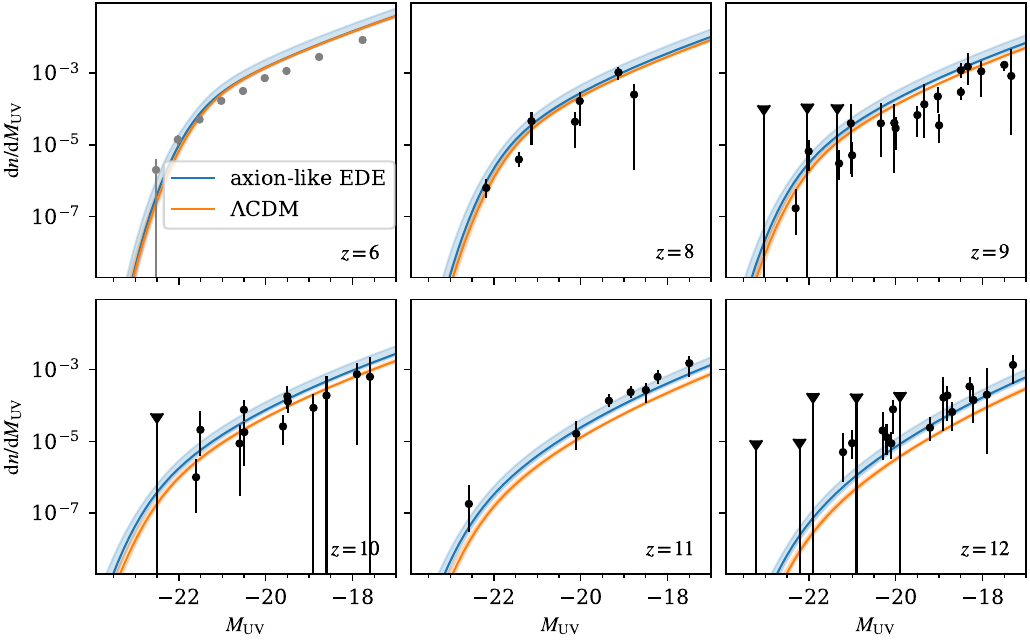}
    \caption{The posterior distribution of the luminosity functions of different models.
    We also plot the JWST data ($z \in [8,\, 12]$) as black error bars.
    In addition, HST measurements of the luminosity function at $z = 6$~\cite{2021AJ....162...47B} (not used in the main text for fitting, but used in Appendix \ref{sec:HST}) are plotted as grey data points in the top left sub-panel.
    All the values have been corrected to the distance - redshift relationship in the $\Lambda$CDM to allow comparison.}
    \label{fig:dndMUV}
\end{figure}

We show the marginalized posterior distribution in \autoref{fig:triangle} and \autoref{tab:posterior}.
We find even when the SH0ES data is not included, a high fraction of EDE $f_\text{EDE} = 0.206^{+0.029}_{-0.026}$ is favored by the JWST data.
The is mainly driven by the JWST data as there is no preference for non-zero $f_\text{EDE}$ if we exclude the JWST data 
(However, see \cite{Smith:2020rxx,Herold:2021ksg} for the discussion of the volume effects).
And it leads to $H_0 = 74.6 \pm 1.2$ km/s/Mpc, which is slightly higher than the SH0ES result \cite{Riess:2021jrx} but is consistent within $0.98 \sigma$.
\footnote{We calculate the tension with $\frac{\abs{\bar{x}_1 - \bar{x}_2}}{\sqrt{\sigma_1^2 + \sigma_2^2}}$.}
Meanwhile, we find $n_s = 1.0071 \pm 0.0078$, which is $1 \sigma$ consistent with $n_s = 1$ (see Refs.\cite{Jiang:2022uyg,Ye:2022efx,Jiang:2022qlj} for the discussion of $n_s=1$ in EDE models.).
At the best-fit points, we find the EDE model fits the data set better with $\Delta \chi^2 = -44.8$ compared to the $\Lambda$CDM model, while we only get $\Delta \chi^2 = -0.3$ if we exclude the JWST data.
Even considering the additional parameters of the EDE models, we still get the improvement of Akaike Information Criterium (AIC) $\Delta \text{AIC} = -38.8$.
In \autoref{fig:dndMUV}, we show the comparison of the posterior distribution of the luminosity functions with the JWST data, where we can find how the EDE model fits the observations at high redshifts better.
However, we find the reduced $\chi^2_\text{JWST}/{\rm dof} = 3.7$ for the EDE model (with dof=62), which means that there may be some unknown uncertainty in the data. 

When the SH0ES data is included, as shown in \autoref{fig:triangle} and \autoref{tab:posterior_wH0}, we get a slightly lower mean value of $f_\text{EDE}$ but with reduced uncertainty $f_\text{EDE} = 0.196^{+0.021}_{-0.018}$.
The Hubble constant $H_0 = 74.11 \pm 0.83$ km/s/Mpc is consistent with the SH0ES result \cite{Riess:2021jrx} with $0.80 \sigma$ and $n_s = 1.0045 \pm 0.0060$ is still within $1 \sigma$ of $n_s = 1$.
Since EDE can help resolve the Hubble tension, the improvement of fitting reaches $\Delta \chi^2 = -77.3$ compared to the $\Lambda$CDM model with the same data set.
This conrespoding to $\Delta \text{AIC} = -71.3$.

\begin{figure}[h]
    \centering
    \includegraphics{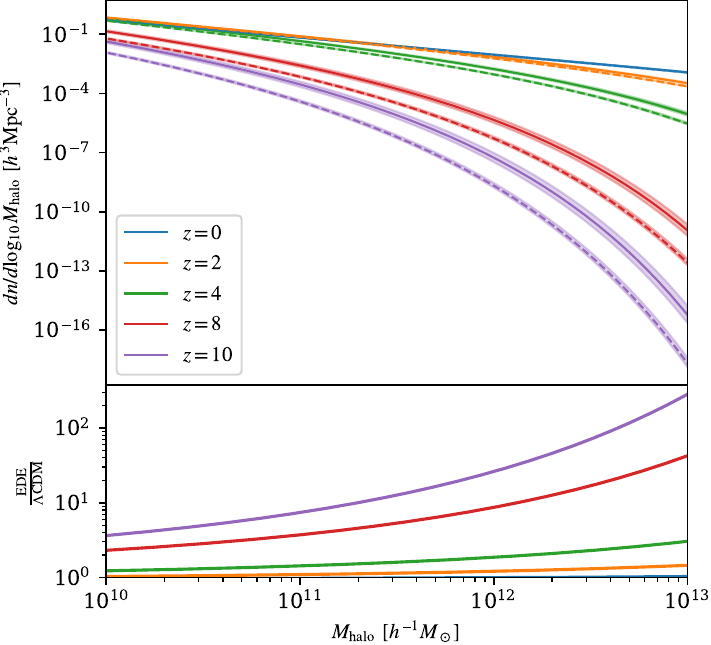}
    \caption{\textbf{Top}: The solid and dashed lines are the halo mass functions of the best-fit points for the EDE and $\Lambda$CDM models for the dataset CMB+BAO+SNe+JWST, respectively. The bands are the 68\% confidence interval of the posterior distribution. \textbf{Bottom}: The relative difference in their halo mass function at the best-fit points points.}
    \label{fig:hmf_EDEvsLCDM_discrete}
\end{figure}

We show in \autoref{fig:hmf_EDEvsLCDM_discrete} the effect of the EDE model on the halo mass function,
which is positively correlated with the observed luminosity function $\dv{n}{M_\text{UV}}$ as shown in \autoref{sec:method_data}.
It is clear that the EDE model leads to more massive halos at high redshifts
which is exactly what the JWST observations have found.
This can be understood by the behavior of the SMT fitting function \autoref{eq:SMT}.
Since the halo mass function decays exponentially ($\sim \exp(-a\nu^2/2)$, which is due to the Gaussianity of the cosmological perturbations) with respect to $\nu = \delta_\text{crit}/\sigma$,
the relative change of the halo mass function with respect to the matter density fluctuation $\sigma$ will be $\sim \sigma^{-3}$,
which means that as the fluctuation $\sigma$ gets smaller, the halo mass function becomes dramatically sensitive to $\sigma$.
Meanwhile,
the matter perturbations are smaller at high redshifts compared to low redshifts since the perturbations are gradually growing (with growth factor $D_+ \sim a$)
and they are smaller on large scales than on small scales (with $\sigma_R \sim k^{-2} T(k_R) \sim \ln(0.12 k_R / k_\text{eq})$).

As a result, massive halos, which correspond to large scales, at high redshifts will be more sensitive to changes in the perturbation with respect to light halos, which correspond to small scales, or halos at low redshifts.
The desired larger matter perturbations compared to $\Lambda$CDM are accomplished in the EDE model by shifts in two parameters.
It is well known that the EDE model leads to a larger $\sigma_8$, which is enhanced by about 6\% for $f_\text{EDE} \sim 0.2$.
On the other hand, for the halo mass ranges we are interested in, a higher $n_s$
in the EDE model also leads to a comparable contribution.
\footnote{
Although a higher $n_s$ can lead to a larger matter power spectrum at smaller scales, which corresponds to lighter halos, it is not comparable to the higher sensitivity of the halo mass function at small scales.
For example, the scales related in our study is $0.02 \text{Mpc} \lesssim R \lesssim 0.2 \text{Mpc}$, the higher $n_s$ in the EDE model will lead to only $2\%$ larger $\sigma_R$ for $R \sim 0.02$ Mpc than $R \sim 0.2$ Mpc, but the relative response of the halo mass function of $R \sim 0.2$ Mpc is $\sim 2^3$ times that of $R \sim 0.02$ Mpc because $\sigma$ is twice smaller there.
}
Eventually, we will obtain a larger halo mass function in the EDE model, especially at high redshifts and large masses.

\section{Conclusion} \label{sec:conclusion}

We tested the axion-like EDE model using the luminosity function measured using JWST.
A six-parameter model is considered to infer the luminosity function from the linear matter power spectrum,
which incorporates the SMT halo mass function \cite{Sheth:1999su},
an AGN-free stellar-to-halo mass ratio \cite{2021ApJ...922...29S},
and a scaling relation between the stellar mass $M_\ast$ and absolute magnitude at the UV band $M_{UV}$
\cite{2021ApJ...922...29S}.
We find the EDE model can fit the data set including JWST with $\Delta \chi^2 = -44.8$ compared to the $\Lambda$CDM model even if the SH0ES data is excluded.
And when the SH0ES data is included, we get an improvement of fitting $\Delta \chi^2 = -71.3$.
Meanwhile, both cases favor high peak fractions of EDE, which lead to $H_0$ consistent with the SH0ES result.
These results reveal the EDE models are not only promising solutions for the Hubble tension, but also for the exceeding massive galaxies at high redshift observed by JWST.

The more massive galaxies at high redshift in the EDE model relative to the $\Lambda$CDM model come from a larger matter density fluctuation at the corresponding scales due to its higher $\sigma_8$ and $n_s$.
On the other hand, the extreme sensitivity of the halo mass function at small fluctuations in the matter power spectrum
leads to the enhancement of the halo mass function concentrating on high redshift and massive halos.

\begin{acknowledgments}
WL and HZ are supported by the National Key R\&D Program of
China grant No. 2022YFF0503400. BH is supported by ``the Fundamental Research Funds for the Central Universities''. 
We acknowledge the use of high performance computing services provided by the International Centre for Theoretical Physics Asia-Pacific cluster and Scientific Computing Center of University of Chinese Academy of Sciences.
\end{acknowledgments}

\appendix

\section{Impact of DESI 2024 BAO measuments} \label{sec:DESI}

\begin{figure}[H]
    \centering
    \includegraphics[width=\textwidth]{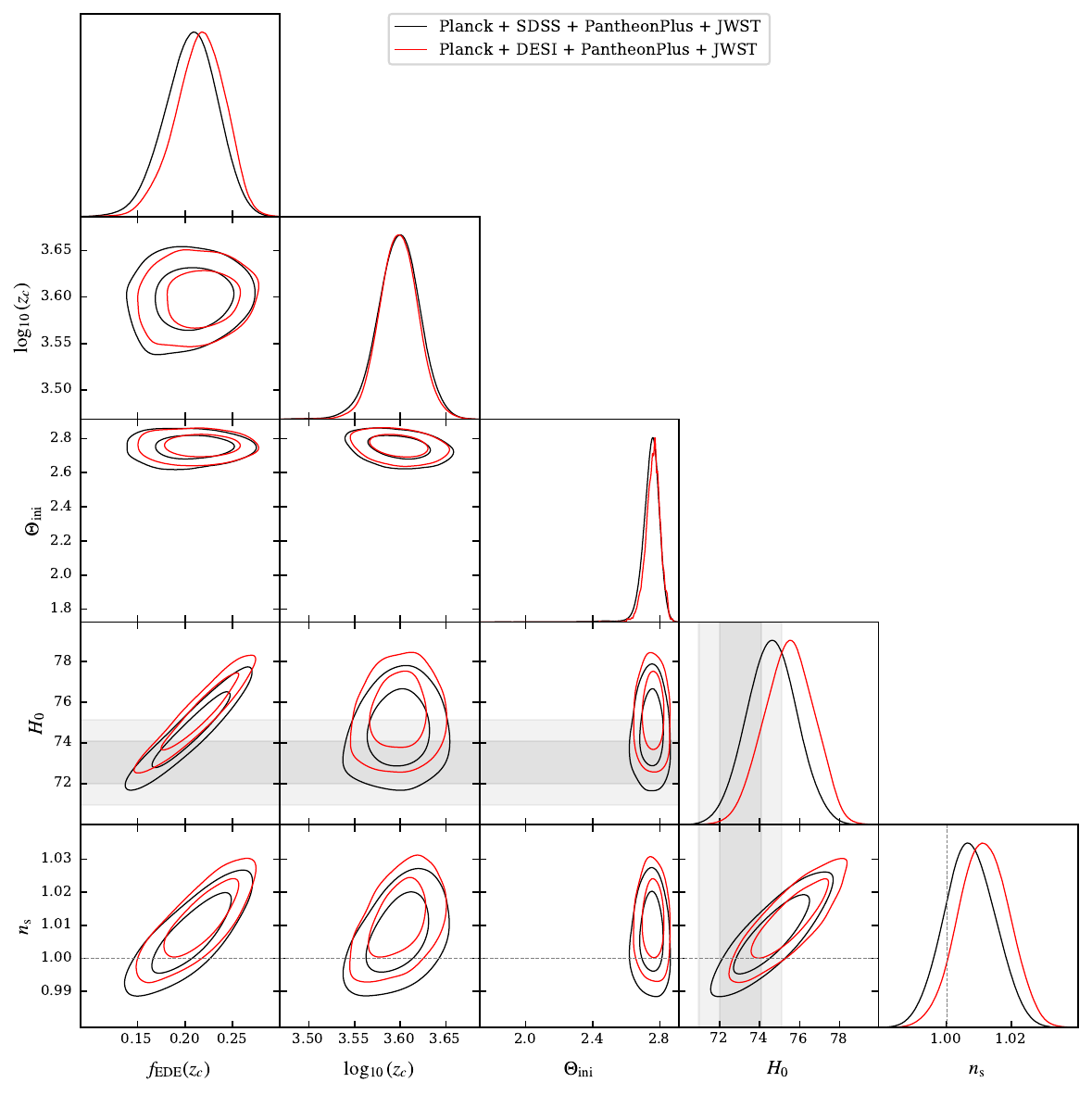}
    \caption{Marginalized posterior distributions for axion-like EDE model with different BAO datasets. The BAO data from 6dF is always included. The contours show the 68\% and 95\% confidence levels. We also show the SH0ES \cite{Riess:2021jrx} constraints on $H_0$ in gray bands.}
    \label{fig:SDSSvsDESI}
\end{figure}

In \autoref{fig:SDSSvsDESI}, we show the constraints on the EDE parameters and some $\Lambda$CDM parameters of replacing the SDSS BAO measurements with DESI results \cite{DESI:2024mwx} (including BGS, LRG, ELG, QSO, Ly-$\alpha$ auto-correlation and its cross-correlation with QSO) by using importance reweighting.
We find a slightly higher $f_\text{EDE}$ and a higher $H_0$.
This can be understood by the lower $\Omega_\text{m}$ preferred by the DESI BAO results.
As CMB constrain $\omega_m = \Omega_\text{m}h^2$, it will lead to a higher $H_0$.
Meanwhile, the DESI BAO results prefer a higher $r_d h^2$ than SDSS
\footnote{See \cite{Jiang:2024xnu,Mukherjee:2024ryz} for consistent conclusions even if late time background evolution is not restricted to the flat $\Lambda$CDM model.}
, which will also lead to higher $H_0$ when combined with the $r_d$ derived from the early Universe.
In both cases, the constraints on $H_0$ from the CMB are weakened, thus enlarging the space allowed for $f_\text{EDE}$.

\section{Impact of HST measurements on the luminosity function at low redshift} \label{sec:HST}

\begin{figure}[H]
    \centering
    \includegraphics[width=\textwidth]{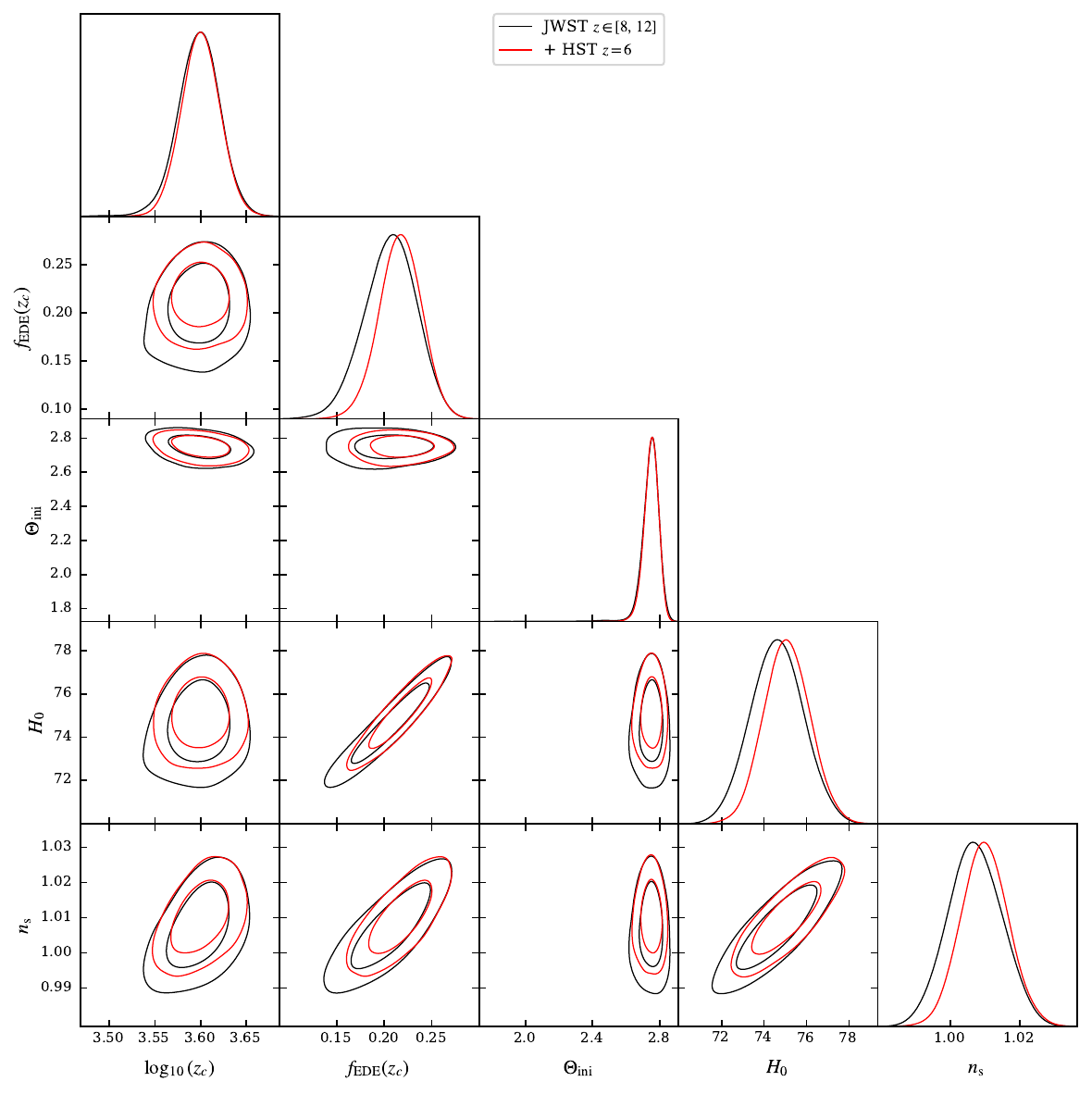}
    \caption{Marginalized posterior distributions for axion-like EDE model for the dataset CMB + BAO + SNe + UV LF.
    The black contours show the constraint with JWST UV LF data, while the red contours show the constraint also includes HST measurements at $z=6$~\cite{2021AJ....162...47B}.
    The contours show the 68\% and 95\% confidence levels..}
    \label{fig:triangle_wz6}
\end{figure}

Here, we investigate the effect of HST measurements on the luminosity function at low redshift.
We neglect the potential stellar feedback-induced redshift dependence of the luminosity function in order to simplify the analysis and perform a joint analysis including constraints on the luminosity function at $z=6$~\cite{2021AJ....162...47B}.
The marginalized posterior distribution is shown in \autoref{fig:triangle_wz6}, where we find a slightly higher $f_\text{EDE}$ if $z=6$ data is included.
This is because the inclusion of data at $z=6$ has increased the demand for the difference in the luminosity functions at high and low redshifts.
The $\Lambda$CDM model has limited degrees of freedom to introduce further redshift dependence, whereas the EDE model can provide a stronger redshift dependence required for the data fitting (see \autoref{fig:hmf_EDEvsLCDM_discrete}).
As a result, EDE is slightly more preferred.

\bibliography{main}
\end{document}